\pdfoutput=1
\documentclass{JINST}
\usepackage{multirow} 

\newcommand{\kr}{\mbox{$\rm ^{83m}Kr$}}
\newcommand{\rb}{\mbox{$\rm ^{83}Rb$}}
\newcommand{\thalf}{\mbox{$T_{1/2}$}}
\newcommand{\be}{\begin{equation}}
\newcommand{\ee}{\end{equation}}
\newcommand{\bea}{\begin{eqnarray}}
\newcommand{\eea}{\end{eqnarray}}

\newcommand{\arxiv}[1]{\href{http://www.arxiv.org/abs/#1}{\tt #1}}

\graphicspath{{figures/}}

\title{Limits on the release of Rb isotopes from a zeolite based \kr\ calibration source for the XENON project}

\author{V.~Hannen$^a$\thanks{Corresponding author.}, E.~Aprile$^b$, F.~Arneodo$^c$, L.~Baudis$^d$, 
M.~Beck$^a$\thanks{Present address: Institut f\"ur Physik, Johannes Gutenberg-Universit\"at Mainz, D-55099 Mainz, Germany}, 
K.~Bokeloh$^a$, A.D.~Ferella$^d$, K.~Giboni$^b$, 
R.F.~Lang$^b$\thanks{Present address: Physics Department, Purdue University, West Lafayette, IN 47907-2036, USA},
O.~Lebeda$^e$, H.-W.~Ortjohann$^a$, M.~Schumann$^d$, A.~Spalek$^e$, D.~Venos$^e$ and C.~Weinheimer$^a$\\
\llap{$^a$}Institut f\"ur Kernphysik, Westf\"alische Wilhelms-Universit\"at M\"unster,
  48149 M\"unster, Germany\\
\llap{$^b$}Physics Department, Columbia University,
  New York, NY 10027, USA\\
\llap{$^c$}Laboratori Nazionali del Gran Sasso,
  67010 Assergi L'Aquila, Italy\\
\llap{$^d$}Physik Institut, University of Z\"urich,
  CH-8057 Z\"urich, Switzerland\\
\llap{$^e$}Nuclear Physics Institute, Acad. Sciences of the Czech Republic, CZ-250 68 Husinec-Rez,\\
  Czech Republic\\

  E-mail: \email{hannen@uni-muenster.de}
}

\abstract{ The isomer \kr\ with its half-life of 1.83~h is an ideal calibration source for a liquid noble gas dark matter experiment like the XENON project. 
However, the risk of contamination of the detector with traces of the much longer lived mother isotop \rb\ ($\thalf = 86.2$~d) has to be ruled out.
In this work the release of \rb\ atoms from a 1.8~MBq \rb\ source embedded in zeolite beads has been investigated.
To do so, a cryogenic trap has been connected to the source for about 10 days, after which it was removed and probed for the strongest \rb\ $\gamma$-rays with an ultra-sensitive Germanium detector.
No signal has been found. The corresponding upper limit on the released \rb\ activity means that the investigated type of source can be used in the XENON project and similar low-background experiments as \kr\ generator without a significant risk of contaminating the detector.
The measurements also allow to set upper limits on the possible release of the isotopes $^{84}$Rb and $^{86}$Rb, 
traces of which were created alongside the production of \rb\ at the Rez cyclotron.}

\keywords{Detector alignment and calibration methods; Noble-liquid detectors}

\begin{document}
\section{Introduction}
The XENON100~\cite{xenon100,xenon100b} experiment aims at a direct detection of weakly interacting dark matter particles in a low-radioactivity liquid xenon (LXe) target. XENON100 will use conversion and Auger electrons as well as $\gamma$-rays from the 
\begin{figure}[b]
\centering
\includegraphics[width=0.7\textwidth]{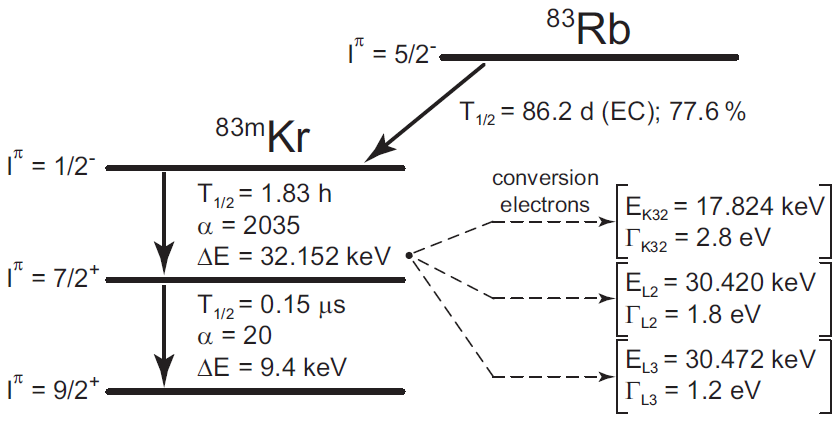}
\caption{Production of the isomeric state \kr\ by electron capture decay of \rb . The isomeric $1/2^-$ state with its half-life of 1.83~h is populated in about 78\% of all \rb\ decay processes and
de-excites by two highly converted $\gamma$-transitions.
Some strong electron conversion lines are also indicated (figure taken from~\cite{thu07}).
\label{fig:kr83m_decay}}
\end{figure}
short-lived isomer \kr\ ($\thalf = 1.83$~h) for calibration of the detector at low energies.
Fig.~\ref{fig:kr83m_decay} shows the decay of the metastable state \kr , which is populated by an electron capture process of \rb .
The conversion electron lines of \kr\ have been extensively investigated at the Mainz Neutrino Mass experiment for calibrations and studies of various systematic uncertainties~\cite{picard92b, aseev, fleischmann1, borschein}.
Also within the KATRIN experiment~\cite{katrin_design_report04}, the K-32 conversion line of \kr\ will be used for several purposes, e.g. for monitoring of the stability of the retarding high voltage at the monitor spectrometer~\cite{kaspar_phd, ostrick_phd, zboril_phd} or for investigations of the properties of the main spectrometer.
The isomer \kr\ is also used to calibrate the time projection chamber (TPC) of the ALICE detector at CERN, which is the worlds largest TPC (about 90~m$^3$)~\cite{alice}.
Several techniques have been developed within the KATRIN collaboration to produce the mother isotope \rb\ at cyclotrons~\cite{rasulbaev,venos05}.

Two features of the isomer \kr , namely the low energy electrons and $\gamma$-rays as well as the short half-life, make it ideal for the calibration of liquid noble gas dark matter detectors. The two subsequent decays with energies of 32.1~keV and 9.4~keV (in general conversion electrons plus fast de-excitation by X-rays or Auger electrons) can be timewise separated by the fast scintillation signal. The total energy of both decays (41.5~keV) can serve for charge calibration. 
However, for a low count rate experiment like XENON100 the risk that traces of the much longer lived mother isotope \rb\ ($\thalf = 86.2$~d) contaminate the detector must be ruled out.
Therefore, any \rb\ source acting as a \kr\ generator needs to be checked for the possible release of \rb\ into the gas phase. Of course the same requirement is true for any application of \rb\ sources in the KATRIN beam line.

The release of \rb\ from a vacuum evaporated source has been investigated at Rez~\cite{diplom_zboril}. The relative sensitivity of this experiment was limited to about 0.2~\%, since the observable was the \rb\ $\gamma$ rate itself, which was checked for its stability (``disappearance'' experiment). 
Another test experiment was performed at the University of Zurich~\cite{man10} with a \rb\ source embedded in zeolite beads.
There, \kr\ signals were searched for in a LXe TPC, after the \kr\ (and possibly \rb ) supply to the TPC was stopped. In principle such an ``appearance'' experiment is more sensitive, but this experiment suffered from a low activity of the \rb\ source (3~kBq). 

In the following we report on a new study of the \rb\ release from zeolite beads, which tries to combine the advantages of the two previous studies (searching for \rb\ $\gamma$-rays, but in an ``appearance'' experiment). We also improved the sensitivity by using a much stronger \rb\ source (1.8~MBq), which was produced at the Nuclear Physics Institute of the Academy of Sciences (AS) of the Czech Republic (CR). \\

The paper is organized as follows: 
In section 2 we describe the \rb\ source and \kr\ generator.
Section 3 describes the test setup at M\"unster, where a cryo trap was exposed to the \rb\ source, whereas in 
section 4 the search for possibly released \rb\ atoms using an ultra-sensitive $\gamma$-detector at Laboratori Nazionali del Gran Sasso (LNGS) is reported on. 
In section 5 we discuss the consequences for the XENON project and summarize the outcome.
\section{$\bf ^{83}$Rb source}
We are investigating \rb\ sources, which were produced at the 
\begin{figure}
\centering
\includegraphics[height=6cm]{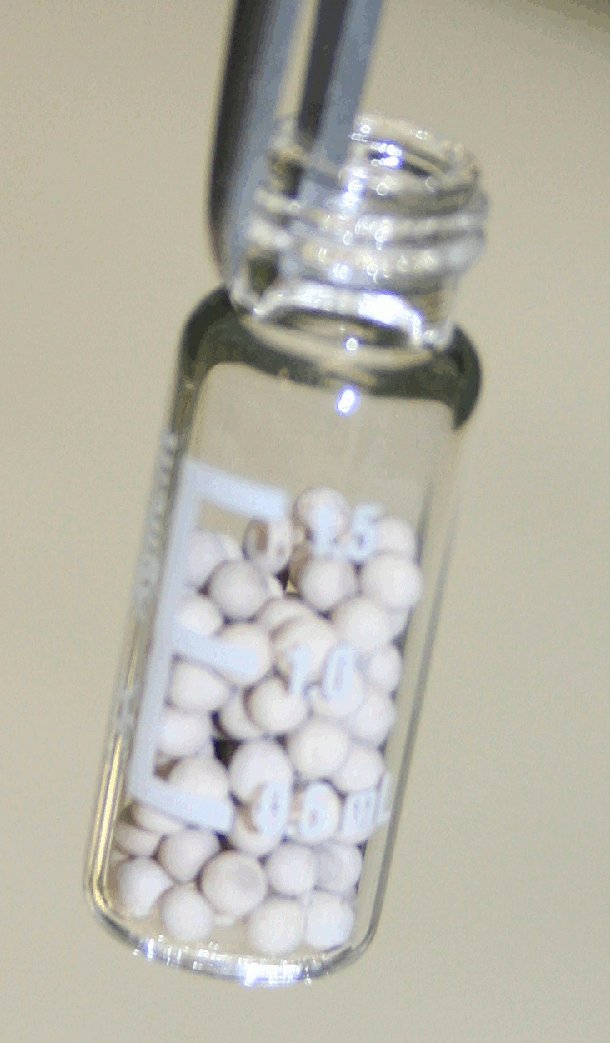} \hspace*{5mm}
\includegraphics[height=6cm]{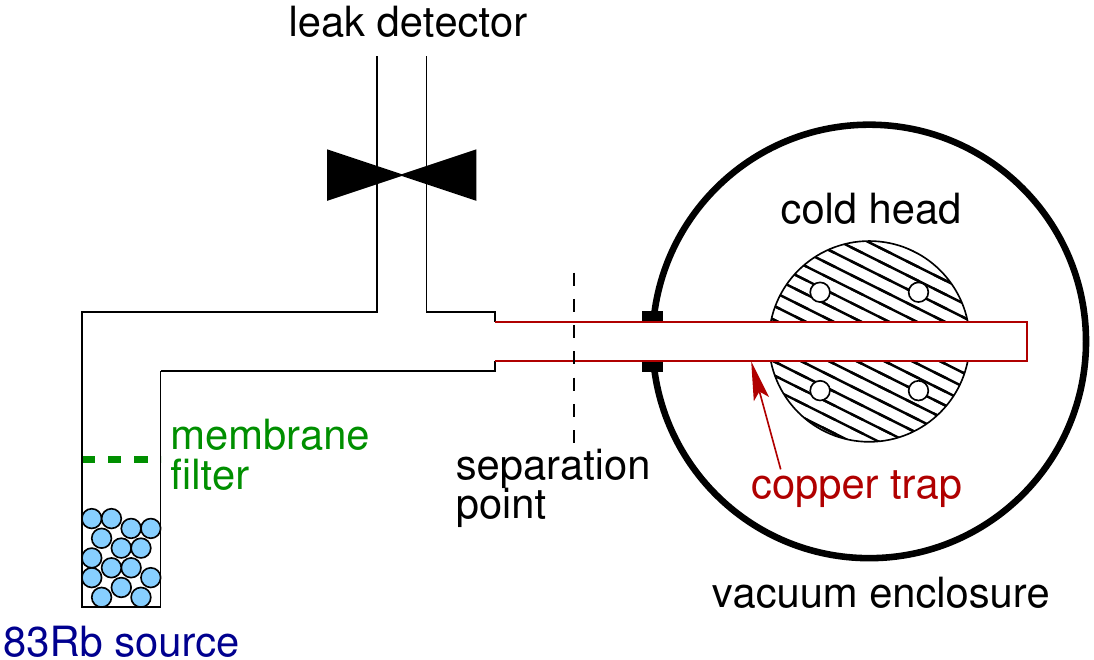}
\caption{Left: zeolite beads of 2~mm diameter tinctured with the \rb\ solution.
Right: Setup of the \rb\ release test.
\label{fig:rb_source}} 
\end{figure}
Nuclear Physics Institute AS CR~\cite{venos05}. 
The radionuclide \rb\ was produced by irradiating a medium-pressurized (7.5~bar) natural krypton gas target by a 26~MeV proton beam of $5-10\,\mu$A. Due to the isotopic composition of natural krypton ($^{78}$Kr 0.0035, $^{80}$Kr, 0.0225, $^{82}$Kr 0.116, $^{83}$Kr 0.115, $^{84}$Kr 0.570, $^{86}$Kr 0.173), the main contribution to the \rb\ production is via the $^{84}$Kr(p,2n)\rb\ reaction. 
There are only two longer-lived radionuclidic impurities, which are produced together with \rb , namely $^{84}$Rb with a half-life of 32.8~d and $^{86}$Rb with 18.6~d half-life. However, these do not emanate any radioactive isotope of krypton. The other rubidium radionuclides have half-lives shorter than 7 hours, so that they decay prior to target processing. 
After irradiation, rubidium isotopes were washed off of the target walls by several portions of deionized water (conductivity $< 0.07\,\mu$S/cm). The volume of the resulting solution was then reduced by evaporation in a quartz beaker under an infrared lamp.

The \rb\ / \kr\ source was prepared by absorbing a portion of the \rb\ solution with the desired activity in zeolite beads~\cite{merck}. Zeolites are crystalline aluminosilicates with very well defined pore sizes ($0.1-2.0$~nm) that act as cation exchanging materials. They are thus very suitable both for trapping \rb\ as counter cation of the aluminosilicate anionic skeleton, and for releasing emanated \kr\ from their porous structure. We used zeolite material with 0.5~nm pores, in the form of beads with 2~mm diameter (see Fig.~\ref{fig:rb_source}, left). The beads were dried for 2~h at 320~$^\circ$C prior to deposition of \rb\ in order to remove all water traces. A small volume of the \rb\ solution was then absorbed by the beads, after which they were dried, first gently at 80~$^\circ$C and finally for 1~h at 320~$^\circ$C. The total activity of the source used for the measurements has been determined via $\gamma$ spectrometry to be 1.8~MBq (31. August 2009).

The beads were put into an aluminum cylinder that was connected to a bent aluminum tube housing a PTFE membrane filter with a pore size of 220~nm~\cite{millipore}. 
The filter allows for easy transport of gaseous \kr\ or \rb\ into the vacuum, but prevents \rb\ in the form of aerosols or zeolite fragments from passing. The bent geometry was used to simplify shielding of the \rb\ source with lead.
\section{$^{\bf 83}$Rb release test}
The idea of the test experiment is -- in a first step -- to collect possibly released \rb\ atoms over a longer period of time in a cryo-trap and -- in a second step -- to disconnect the trap from the \rb\ source and to perform a very sensitive $\gamma$ spectroscopy measurement on the trap to look for lines from a potential \rb\ contamination in the spectrum.
Therefore, the trap has to be made from material with very low specific radioactivity and needs to be small enough to fit into the $\gamma$-spectroscopy setup at LNGS.

\begin{figure}
\centering
\includegraphics[height=7cm]{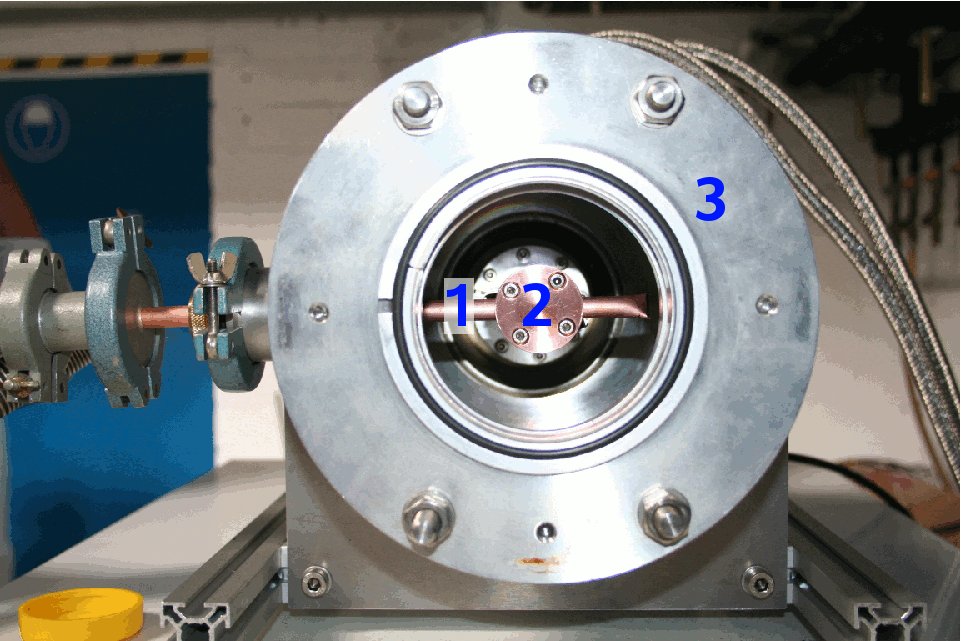}
\includegraphics[height=7cm]{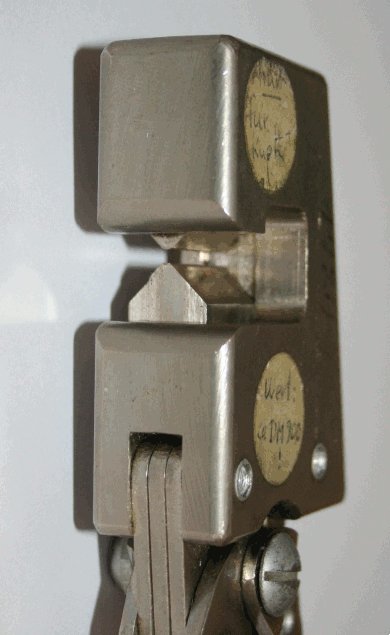}
\caption{Left: copper tube (1) screwed onto the cold-head (2) sitting in the insulating vacuum chamber (3). Right: pliers used for leak tight cold-welding of the copper tube.
\label{fig:coldhead}} 
\end{figure}

To fulfill these requirements, we decided to use a small cryogenic trap consisting of a low-activity OFHC copper tube 
kept at a temperature below 30~K by a two stage cryo-cooler of Gifford-McMahon type. In order to allow for the disconnection of the copper tube without a voluminous and possibly radioactive valve we used a special cold-welding technique to cut and to seal the tube leak-tight with the help of 
%
%
special pliers (see Fig.~\ref{fig:coldhead}, right)\footnote{This technique has been used successfully at the Mainz Neutrino Mass experiment to separate tritium containing parts.}.
The copper tube was cleaned with citric acid and ultra-pure water in an ultrasonic bath. One side of the copper tube was sealed by cold-welding, the other side was connected to the \rb\ source via a t-connector (see Fig.~\ref{fig:rb_source} right).
The third leg of the t-piece was connected via a valve to a leak detector which was used for pump-down of the trap volume and for a check of the leak-tightness of the setup including the cold-welded copper tube. After pump-down to a pressure of about $10^{-2}$~mbar, both the \rb\ source and the copper tube were separated from the leak detector by closing the valve.
The copper tube was mounted within an insulating vacuum vessel onto a cold head 
(see Fig.~\ref{fig:coldhead}, left) and cooled down to 28~K, which is by far enough to efficiently trap rubidium.

%
%
We collected potentially released \rb\ with this setup for 268~h in August 2009.
At the end of the run 
the copper tube was cut and cold-welded on the room-temperature side. 
While still being cold, the tube was dismounted from the cold-head.
The tube was subsequently shipped to LNGS to be screened for possible traces of \rb, $^{84}$Rb and $^{86}$Rb isotopes.
%
%
\section{Counting results}
\begin{figure}
\centering
\includegraphics[width=0.9\textwidth]{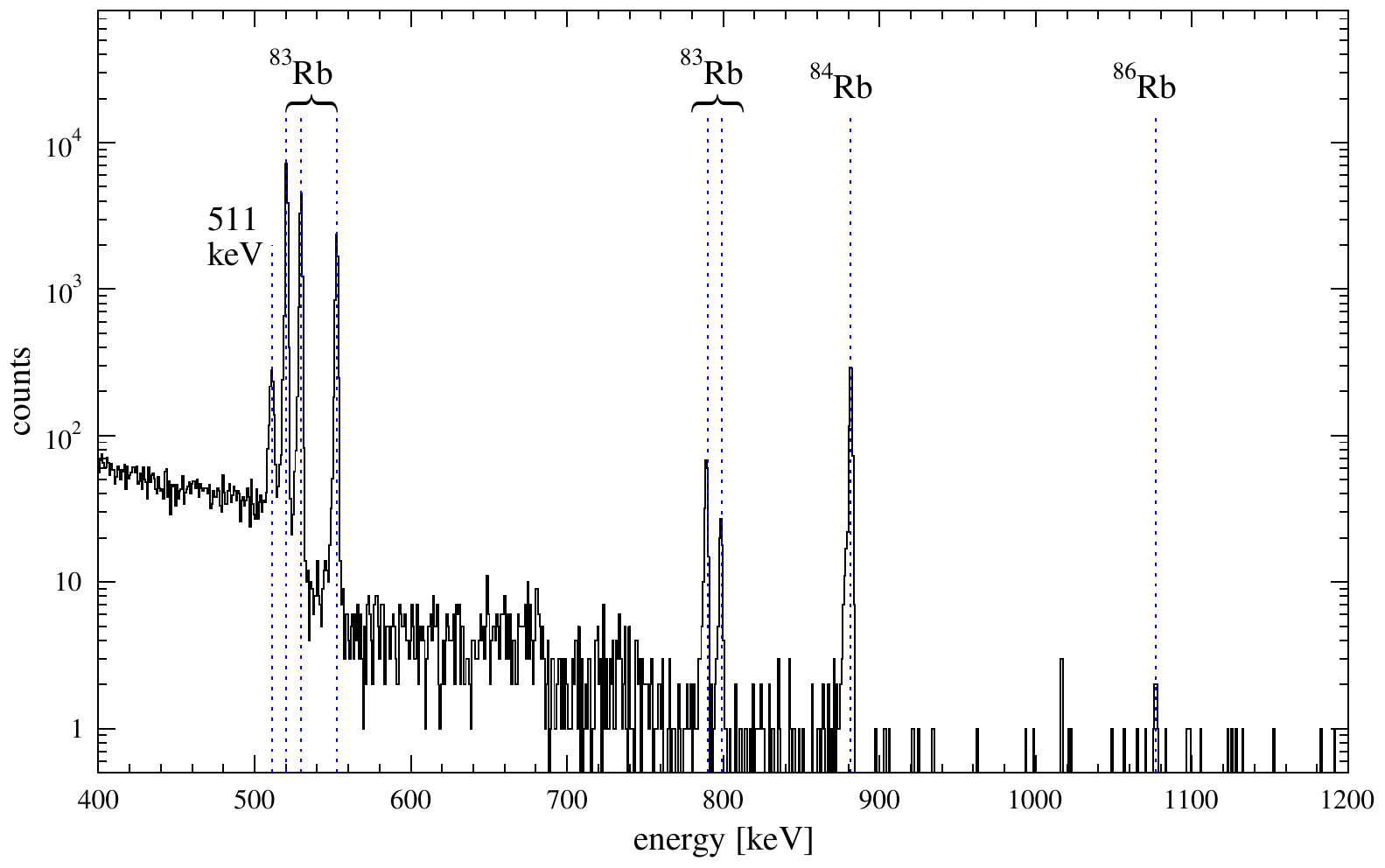}
\caption{Part of the energy spectrum from a \rb\ source embedded in zeolite measured with the Gator detector for calibration purposes.
\label{fig:rb_label}} 
\end{figure}

The copper tube exposed to the \rb~source was $\gamma$-ray counted (screened) in the Gator low background germanium detector facility~\cite{gator} at LNGS. In order to determine the exact energy scale, a calibration run with a \rb~source was performed prior to the measurement of the sample (see Fig.~\ref{fig:rb_label}). 
By comparison to the Gator background spectrum~\cite{gator}, 
it was verified that no background lines are present in the region of interest (ROI).
A reference tube, made of the same material but not exposed to the \rb\ source, was measured in the
same facility in order to verify that the tube itself is not contaminated with $\gamma$-lines in
the ROI for Rb counting. In this measurement, that lasted 10.53 days, no counts above the background of the spectrometer were found in the ROI.

\begin{figure}
\centering
\includegraphics[width=1.0\textwidth]{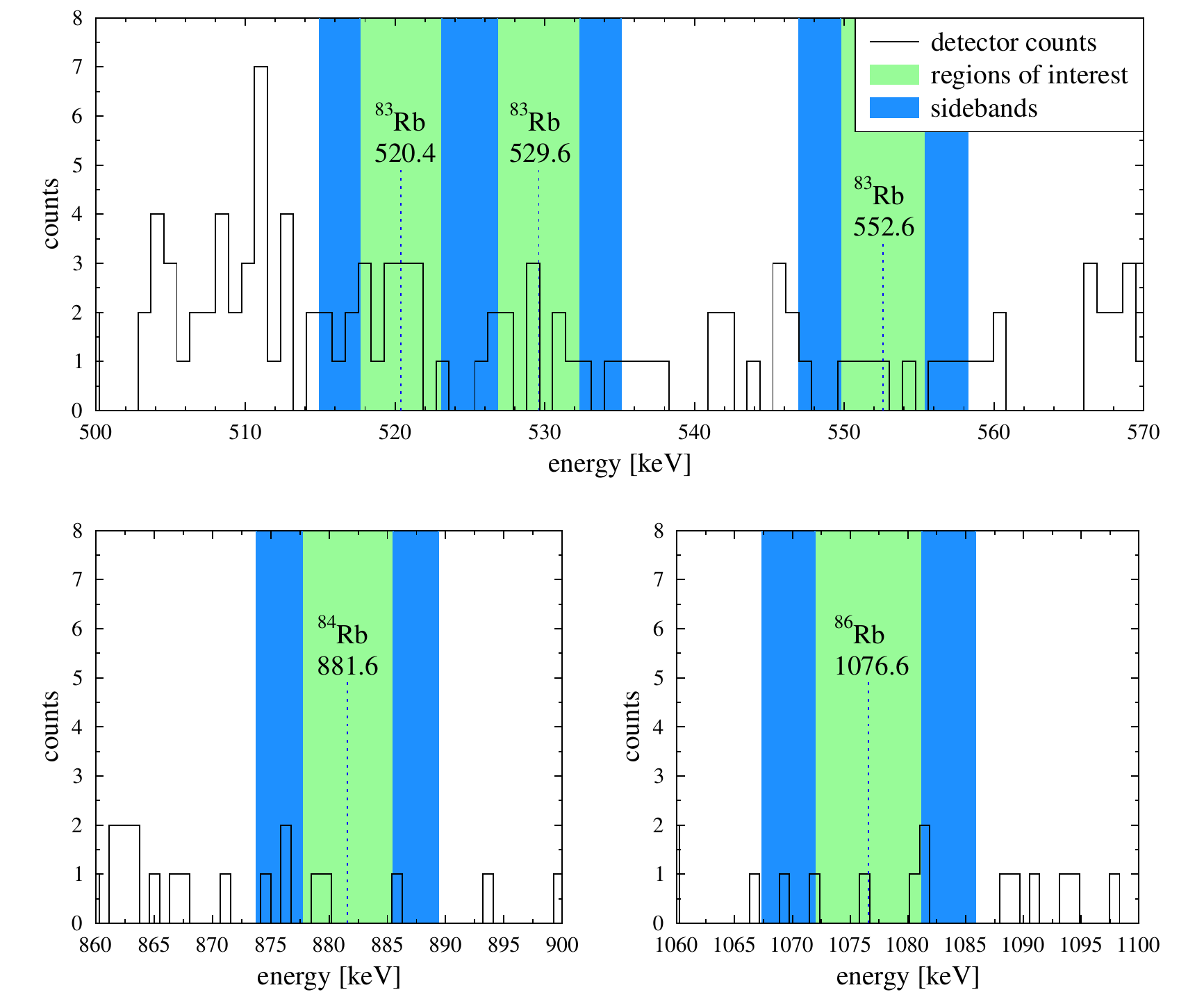}
\caption{Measured spectrum of the copper tube after exposure to the \rb~source, zoomed into the regions of interest.
The $\pm 3\sigma$ areas around the $\gamma$-lines are marked in green, while the sidebands of the same total widths are indicated in blue. \label{fig:rb_analysis}} 
\end{figure}

The copper tube that had been exposed to the \rb~source at M\"unster, was measured for 18.94~days and the resulting $\gamma$-ray spectrum was analyzed in order to determine the average sample activity $A_{\rm ave}$ during the measurement.
The activity (or upper activity limit) for $\gamma$-lines of Rb isotopes that are contained in the zeolite source and could therefore possibly be released into the copper tube during the test, was calculated from the observed number of counts $S_{raw}$ in a $\pm3\sigma$ region around the expected peak position, after subtracting the Compton background from higher energetic lines in the spectrum. 
The Compton background $B_C$ is determined for each peak from two $3\sigma$ wide sidebands above and below the signal region. Because the analysis allowed for fractional bins to be taken into account, the resulting count numbers can differ from integers. 
The regions of interest for the analysis are shown in Fig.~\ref{fig:rb_analysis}, where the peak areas are marked green
and the sidebands used to determine the background are indicated in blue.
The resulting numbers of counts in the peak windows $S_{raw}$ and for the Compton 
\begin{table}
\centering
\caption{Results for the $\gamma$ lines under investigation from \rb,  $^{84}$Rb and $^{86}$Rb (energies and half-lives from Ref.~\cite{toi}). In the table, $S_{raw}$ denotes uncorrected counts in the peak areas, $B_C$ is the Compton background from the sidebands, $L_d$ are detection limits resulting from the observed background counts, $U_L$ are upper limits derived from net signal counts and detection limits, BR are the branching ratios of the transitions, $\epsilon$ the detection efficiencies, $A_{\rm ave}$ the average observed activities within the measurement period and $A_0$ the activities of the tube right after separation from the \rb\ source. Upper limits are specified at 95\% CL.}

\vspace{2mm}

\begin{tabular}{|c|c|c|c|c|c|c|c|c|c|c|c|} \hline
isotope   & $T_{1/2}$ & energy & $\sigma$ & $S_{raw}$& $B_C$ & $L_d$ & $U_L$  & BR   & $\epsilon$ & $A_{\rm ave}$ & $A_0$  \\ \hline
          & [d]       & [keV]  & [keV]    & [cts]    & [cts] & [cts] & [cts]  & [\%] & [\%]       & [mBq]       & [mBq]  \\ \hline \hline
\rb       & 86.2      &  520.4 & 0.91     & 13.0     &  6.6  & 16.3  & < 22.7 & 44.7 & 4.00       & < 0.78      & < 0.89 \\ 
          &           &  529.6 & 0.92     &  8.6     &  4.7  & 14.6  & < 18.5 & 29.3 & 3.97       & < 0.97      & < 1.11 \\ 
          &           &  552.6 & 0.94     &  4.8     &  4.3  & 14.2  & < 14.7 & 16.0 & 3.89       & < 1.44      & < 1.65 \\ \cline{3-12} 
          &           &  comb. &          & 26.4     & 15.6  & 22.5  & < 33.3 & 90.0 & 3.97       & < 0.57      & < 0.65 \\ \hline\hline
$^{84}$Rb & 32.8      &  881.6 & 1.30     &  2.1     &  3.9  & 13.8  & < 13.8 & 69.0 & 2.71       & < 0.45      & < 0.64 \\ \hline\hline
$^{86}$Rb & 18.6      & 1076.6 & 1.54     &  2.9     &  3.1  & 13.0  & < 13.0 &  8.6 & 1.99       & < 4.65      & < 8.63 \\ \hline
\end{tabular}
\label{tab:results}
\end{table}
background $B_C$ are given in Tab.~\ref{tab:results}.

To calculate sensitivities and upper limits for the isotopes of interest, we follow the approach of Hurtgen et al.~\cite{hurtgen}: The detection limit $L_d$ for finding a signal at 95\% confidence level (CL) on top of a background $B_C$ is given by
\begin{equation}
 L_d = 2.86 + 4.78\cdot\sqrt{B_C + 1.36} \; ,
\end{equation}
where the coefficients are calculated from a coverage factor of 1.69 for a one-tailed probability distribution.
This detection limit has to be compared to the net number of signal counts given by 
\begin{equation}
  S_{net} = S_{raw} - B_C \; .
\end{equation}
In case we have $S_{net} < 0$ an upper limit $U_L$ corresponding to the detection limit $L_d$ is quoted. For $0 < S_{net} < L_d$ an upper limit of $S_{net} + L_d$ is given. An activity corresponding to the observed number of net counts is reported for the case $S_{net} > L_d$.
As apparent from Tab.~\ref{tab:results}, there are no significant count excesses above background, and,
therefore, only upper limits can be given
\footnote{Using alternative statistical methods, like the approach described by Feldman and Cousins in Ref.~\cite{Fel98}, it is possible to extract more stringent upper limits than with the more conservative approach of Hurtgen et al.~\cite{hurtgen}. However, this method requires the \emph{mean} background to be known, whereas we only have one background estimate per peak. 
If we just assume that our count numbers have Gaussian uncertainties -- which is not fully correct for small numbers -- we can extract upper limits at 95\% C.L. by the Feldman Cousins method (table X in Ref.~\cite{Fel98}), which are factors 1.4, 4.5 and 2.9 times lower than our limits for the isotopes \rb, $^{84}$Rb and $^{86}$Rb, respectively. The larger factors reflect our negative net signal for $^{84}$Rb and $^{86}$Rb.
%
}.

\begin{figure}[h]
\centering
\includegraphics[width=0.8\textwidth]{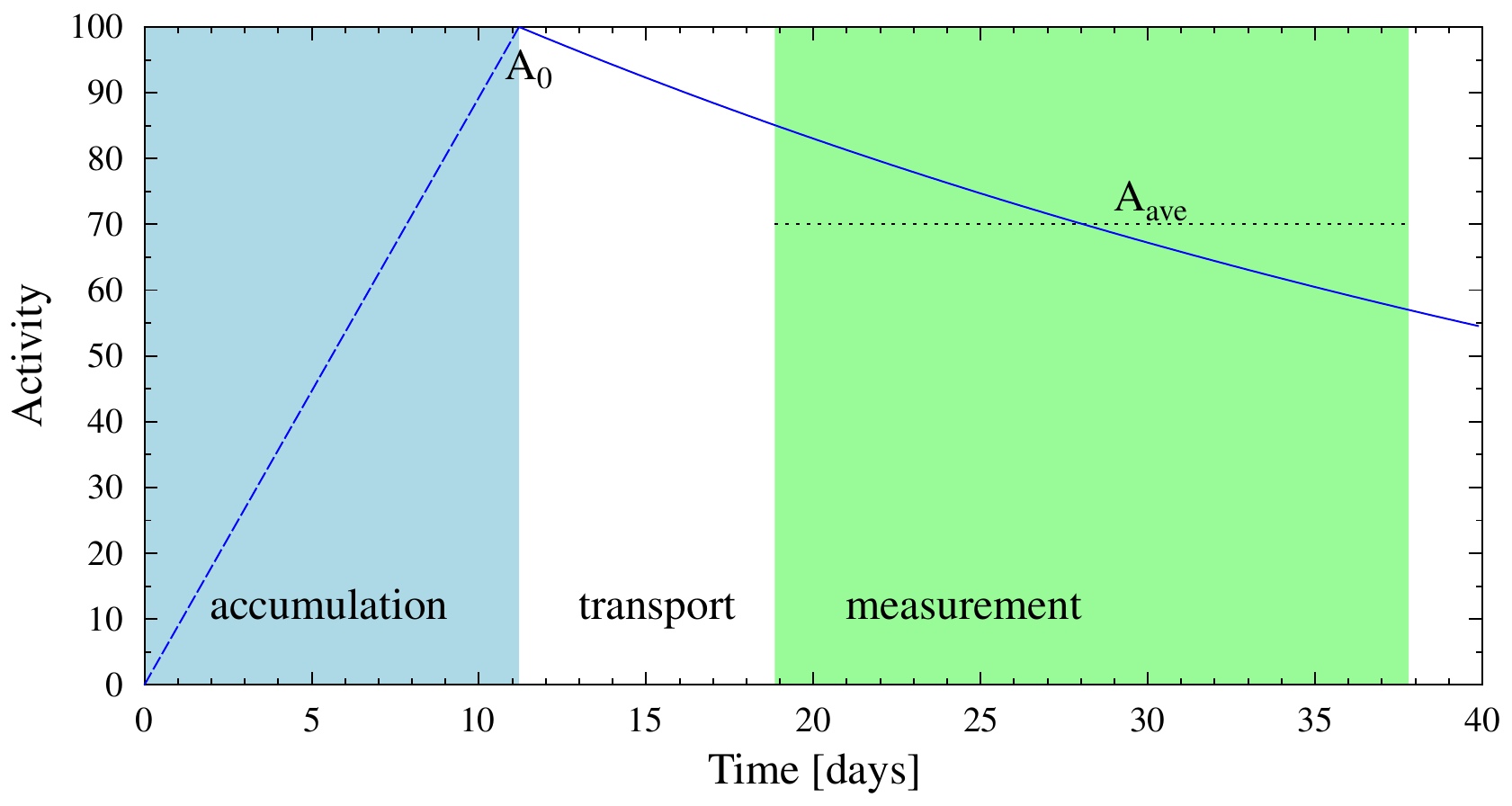}  
\caption{Build up and decay of a possible $^{84}$Rb contamination ($T_{1/2} = 32.8$~d) in the sample during the test.
\label{fig:average}} 
\end{figure}

In order to transfer these numbers into upper limits on the observed activity in Becquerel, we take into account the branching ratios (BR) of the transitions as well as the detection efficiencies $\epsilon$ for the different peak energies.
The detection efficiencies for this analysis were obtained from a Monte Carlo simulation, using the Geant4 package~\cite{geant4}, where the complete sample and detector geometry was modeled. 
With this information we can then determine activity limits $A_{\rm ave}$ that result from the average activity of the sample observed during the Gator measurement. In order to calculate upper limits $A_0$ of the activity of the sample right after it has been separated from the \rb~source (corresponding to the activity that would be introduced into the dark matter detector), we need to correct for the decay of Rb isotopes during the shipping time (7.65~days) to LNGS and during the measurement time of 18.94~days (see Fig.~\ref{fig:average}), taking into account the half-lifes of the respective isotopes. 
The resulting upper limits on the contamination of the sample with Rb isotopes right after it had been in contact with the zeolite source for 11.17~days vary between $A_0<0.65$~mBq for a combined analysis of the three \rb~lines and $A_0<8.63$~mBq for $^{86}$Rb
(see Tab.~\ref{tab:results}).

To verify that the sealed copper tube that was examined at LNGS was hermetically sealed throughout the measurement, it was sent back to M\"unster, cut in the middle and both ends were checked with a leak tester. The cold welds at the ends of the tube were found to be still leak tight.
\section{Discussion and conclusions}
Ref.~\cite{man10} describes a release test of a 3~kBq \rb\ source into the X\"urich LXe TPC with an exposure time of 150~hours. The measurement led to an upper limit of $r_\mathrm{X\ddot urich} \leq 160~\mu$Bq/h at 90\% CL (corresponding to $r_\mathrm{X\ddot urich} \leq 191~\mu$Bq/h at 95\% CL) for the possible \rb\ contamination of the TPC normalized to the exposure time. In this test the \rb\ source was directly connected to the xenon gas line towards the X\"urich detector through a filter.
To derive an upper limit on the actual background rate caused by a potential release of \rb\ from a zeolite source, the authors have performed a Monte Carlo simulation for a 300~kg LXe detector with 100~kg fiducial mass that is assumed to be connected for 10~h to a 3~kBq \rb\ source. The simulation resulted in a background contribution of less than $67\,\mu$DRU (1~DRU~$\equiv 1$~event/kg/day/keV) in the WIMP search region,
without applying discriminating cuts on the signal characteristics. The corresponding background rate from natural radioactivity of the detector materials has been estimated to be of the order of 1~mHz, well above the extracted upper limit on the \rb\ activity.
For upcoming dark matter experiments employing several tons of liquid xenon~\cite{xenon1t,darwin} one would, however, prefer to use stronger 
calibration sources that require more stringent limits on a possible release of \rb\ into the detector volume.

In this work we find an upper limit for a possible \rb\ contamination of a cold trap that had directly been connected for 268~hours to a \rb\ zeolite source of $r_\mathrm{this work} \leq 2.4~\mu$Bq/h (at 95\% C.L.). Since we used a 600 times stronger \rb\ source (1.8~MBq at the end of the exposure interval), we achieve a strong improvement in the sensitivity on the \rb\ release per source activity. The improvement on the earlier measurement can be quantified using the factor $\alpha$ given by 
\begin{equation}
  \alpha = \frac{1.8~{\rm MBq}}{3~{\rm kBq}} \cdot \frac{191~\mu{\rm Bq/h}}{2.4~\mu{\rm Bq/h}} \approx 4.8\cdot 10^4
\end{equation}
To estimate an upper limit on the \rb\ induced background rate $B_{\rm 83Rb}$ in the relevant energy region for a hypothetical LXe detector with a ratio $f$ of fiducial mass to total mass that is connected for $t_e$ hours to a \rb\ source of activity $A_s$, 
we scale the Monte Carlo result from Ref.~\cite{man10} to obtain
\begin{equation}
  B_{\rm 83Rb} ~\leq~ f \cdot t_e \cdot A_s \cdot \frac{67\,\mu{\rm DRU}}{\alpha \cdot 3~{\rm kBq} \cdot 10~{\rm h}} \cdot 
                      \frac{300~{\rm kg}}{100~{\rm kg}} 
                ~=~ f \cdot t_e \cdot A_s ~\cdot~1.4\cdot 10^{-4} ~\frac{\mu{\rm DRU}}{{\rm kBq~h}} 
\end{equation}
Our limit only holds as long as the zeolite beads containing the \rb\ are placed in a vacuum environment, because
\begin{figure}
\centering
\includegraphics[width=1.0\textwidth]{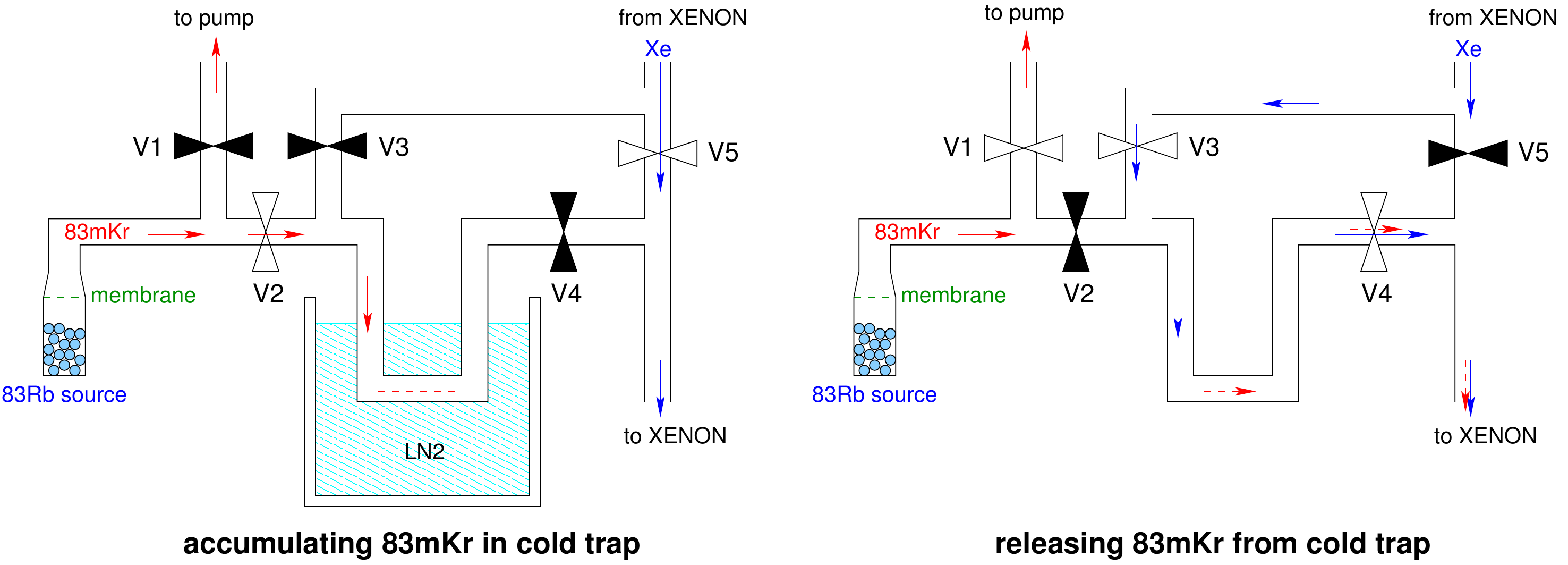}  
\caption{Possible setup for calibrating a XENON detector with \kr\ without any risk of releasing \rb\ into the XENON setup.
\label{fig:rb83_application}} 
\end{figure}
we cannot deduce the \rb\ release rate of the zeolite spheres in an environment of gaseous or liquid xenon from our measurements.
Therefore, a safe way for the calibration of a LXe detector would be to use a setup as is shown in 
Fig.~\ref{fig:rb83_application}.
There, a cold trap is connected to the \rb\ source and the xenon gas lines by a tubing system with several valves. 
For accumulation of \kr\ in the trap the valve V2 to the \rb\ source is opened, while the valves V3 and V4 to the xenon lines are closed. The trap is cooled, e.g. using liquid nitrogen, to efficiently collect \kr\ atoms. 
For calibration of the detector, the valve V2 to the \rb\ source needs to be closed before the copper tube is warmed up to release the \kr\ (and to avoid condensing Xe). Subsequently, the valves V3 and V4 to the xenon gas lines are opened to flush out the collected \kr .\\

Summarizing, we can state that the studied \rb\ source embedded in zeolite beads does not release \rb\, nor the isotopes $^{84}$Rb and $^{86}$Rb, at a detectable level. We have improved this limit by a factor $4.8\cdot 10^4$ compared to previous investigations. Therefore, this source can be used within the XENON project for calibrations with \kr\ without any significant risk of contamination. 

The result also holds for the KATRIN experiment, where conversion electrons from \kr\ will be used for calibration of the spectrometers. A requirement on the allowed background from a potential release of \rb\ can be deduced from the fact that an upper limit of 1~mHz has been set on the background rate produced by residual tritium molecules in the main spectrometer~\cite{katrin_design_report04}. 
With an activity $\leq 2.4~\mu$Bq per hour of exposure to a 1.8~MBq \rb\ source, 
one would be below this limit of 1~mBq with any exposure less than 2.5 weeks, even without taking into account that the KATRIN spectrometer is equipped with a powerful pumping system that would also remove Rb from the spectrometer volume.
\acknowledgments
The work of the M\"unster group on XENON100 is supported from 2011 on by the Deutsche Forschungsgemeinschaft.
The group from Z\"urich is supported by the SNF Grant 200020-138225 and by the University of Z\"urich.
Three of us (O.L., A.S. and D.V.) acknowledge the support of the Czech Ministry of Education - grant LC07050.

{}
\end{document}